# High Entropy Engineering of Magnetic Kagome Lattice (Gd,Tb,Dy,Ho,Er)Mn$_6$Sn$_6$

Wenhao Liu[1,2], Nikhil Uday Dhale[1], Youzhe Chen[2,3], Pramanand Joshi[4], Zixin Zhai[1], Xiqu Wang[5], Ping Liu[4], Robert J. Birgeneau[2,3], Boris Maiorov[6], Christopher A. Mizzi[6], Bing Lv[1,*]

1. Department of Physics, the University of Texas at Dallas, Richardson, TX 75080, USA
2. Department of Physics, University of California, Berkeley, CA, 94720, USA
3. Material Sciences Division, Lawrence Berkeley National Lab, Berkeley, California 94720, USA
4. Department of Physics, the University of Texas at Arlington, Arlington, TX 76019, USA
5. Department of Chemistry, University of Houston, Houston, TX 77204, USA
6. National High Magnetic Field Laboratory, Los Alamos National Laboratory, Los Alamos, NM 87545, USA

* to whom the correspondence to be addressed: blv@utdallas.edu

**The magnetic kagome lattice compound RMn$_6$Sn$_6$ (R=rare earth) is an emerging platform to exploit the interplay between magnetism and topological electronic states where a variety of exciting findings such as flat bands, Dirac points as well as the dramatic dependence of magnetic order on the rare-earth element have been reported. High entropy through rare earth alloying, on the other hand, provides another knob to control over the physical properties in this system. Here, by the marriage of high entropy and the magnetic kagome lattice, we obtain (Gd,Tb,Dy,Ho,Er)Mn$_6$Sn$_6$ single crystals and systematically investigate their magnetic and transport properties. Different from the parent phases, the high entropy 166 material displays multiple novel magnetic transitions induced by temperature and external magnetic fields. Furthermore, linear magnetoresistance persisting up to 20 T has been revealed at 4 K. The intrinsic nontrivial band topology also survives in the high entropy form, as evidenced by the intrinsic anomalous Hall effect. Our results highlight high entropy as a powerful approach for tuning the interplay of charge, spin and lattice degree of freedom in magnetic topological materials.**

The kagome lattice has received great attention recently due to the novel physics and the associated interesting properties arising from the interplay between electron correlations and lattice geometry. A typical kagome lattice consists of corner shared triangles, forming a distinctive honeycomb-like structure.[1] This unique geometry has great potential to realize novel band structures including flat bands, linear band crossings forming Dirac/Weyl points and van Hove singularities, which harbor important correlated electronic states and lead to rich interesting phenomena such as unconventional superconductivity[2], dissipationless spin moment locked charge transport and large quantum anomalous Hall effects[3,4]. Introducing magnetic ions into a geometrically frustrated kagome lattice provides another tool to tune these correlated electronic states. A prime example is hexagonal RMn$_6$X$_6$ (R166), where R represents transition metal elements like Sc, Y, and rare earth elements La-Lu. These compounds usually consist of two sublattices: a Mn-Sn layer with a kagome lattice of Mn and a triangular R-Sn layer. The Mn kagome layer shows strong ferromagnetic in-plane coupling[5], which results in magnetic order above room-temperature.

On the other hand, Mn and the rare earth elements are generally antiferromagnetically coupled[6,7]. Competition between these ferromagnetic and antiferromagnetic couplings results in rich magnetic phases. For example, collinear ferrimagnetic order (FIM) in (Gd-Ho)$Mn_6Sn_6$ emerges when strong antiferromagnetic coupling between the Gd-Ho and Mn sublattices dominate. For non-magnetic rare earth elements Sc and Y, the ferromagnetic coupling between Mn is strongest and their R166 phases show a spiral-magnetic structure in which the ferromagnetic magnetic order rotates among adjacent Mn layers[7–9]. The coupling between Mn and Er is at the boundary between the formation of FIM in (Gd-Ho)$Mn_6Sn_6$ and spiral magnetism in (Y, Sc)$Mn_6Sn_6$. As a result, Er$Mn_6Sn_6$ shows an interesting triple-spiral magnetic ordering[10,11] and topological Hall effect[7,12,13]. As a whole, R166 shows unique temperature and field-driven magnetic instabilities, opening new avenues in topological and magnetic state control and switching.[14]

High entropy is another concept that received great attention in the past decade. It is an alloying strategy that yields a single phase random distribution of solid solution with five or more elements.[1,15,16] The formation of high entropy materials relies on the configuration entropy contributing to the free energy sufficiently to overcome the formation enthalpy, resulting in a highly disordered yet chemically homogeneous system. This high degree of homogeneous in high entropy materials plays a crucial role in including various material properties like mechanical, electronical and magnetic characters. Moreover, it provides a vast compositional space with diverse combinations of composition, charge and spin, enabling the exploration of much wider range of novel correlated phenomena.[1,15] Motivated by the potential advantages of this strategy and the rich magnetic structures specified by individual rare-earth elements in the R166 system, we synthesized and characterized a new high entropy kagome lattice by incorporating five rare earth elements – Gd, Tb, Dy, Ho, Er – on the same crystallographic site, forming a high entropy kagome phase in ($Gd_{0.21}Tb_{0.22}Dy_{0.22}Ho_{0.19}Er_{0.16}$) $Mn_6Sn_6$, denoted as HE166 in the main text. Large-size single crystals have been successfully grown via Sn flux method. A cascade of new phenomena emerges in this high-entropy kagome system that is not found in the parent compounds. Magnetic measurements reveal multiple magnetic transitions including ferrimagnetic transitions and spin reorientations. The application of an external magnetic field additionally induces multiple field-dependent transitions including metamagnetic states and possibly chiral magnetic order. Correlated with magnetization, our electrical transport results show a first-order phase transition around 205 K. Magneto-transport measurements exhibit a linear, non-saturating magnetoresistance persisting up to 20 T, potentially linked to the band topology and the high entropy effects. Hall measurements further reveal the intrinsic Hall effect and non-zero internal Hall conductivity, indicating the survival of Chern gapped Dirac Fermions that exist in the parent phases. These findings demonstrate the power of high-entropy engineering in tailoring the interplay between charge, spin, and lattice degrees of freedom in magnetic topological materials.

The crystal structure of HE166 with five rare-earth elements was determined using X-ray diffractometer on the single crystals. Fig. 1a presents the procession image of (*0kl*) zone, constructed using a set of 2404 measured φ scan frames for 123 K and 2097 for 300 K. The sharp Bragg peaks, free from heavy distortions like ring shapes or blurry spots, shows the high crystallinity in our single crystals, which is further confirmed by the good refinement results showing in the Tabel 1. X-ray diffraction and related refinements indicate the HE166 crystal with five rare earth elements has the same crystal structure as their parent phases where the rare earth elements are randomly distributed on the Wyckoff position 1a site. Moreover, we do not observe a structure change between 123 K and 300 K within the resolution of X-ray diffraction equipment, as shown in Fig. 1a and Table 1, excluding a structure change at the magnetic phase transitions which we will discuss later. The homogeneous distribution of rare earth elements is further confirmed with energy-dispersive X-ray spectroscopy (Fig. 1c), which suggests the successful formation of high entropy. The measured composition of HE 166 from SEM-EDX is ($Gd_{0.21}Tb_{0.22}Dy_{0.22}Ho_{0.19}Er_{0.16}$) $Mn_6Sn_6$, which is rather close but slightly different from the nominal evenly distributed (20% for each rare earth) nomination composition starting the synthesis. It is interesting to note that, the HE 166 crystals are directly grown from elements without obvious other phases formation during the synthesis, while during single rare earth 166 phase synthesis, minor impurities such as RSn$_2$ with different crystal morphology is observed

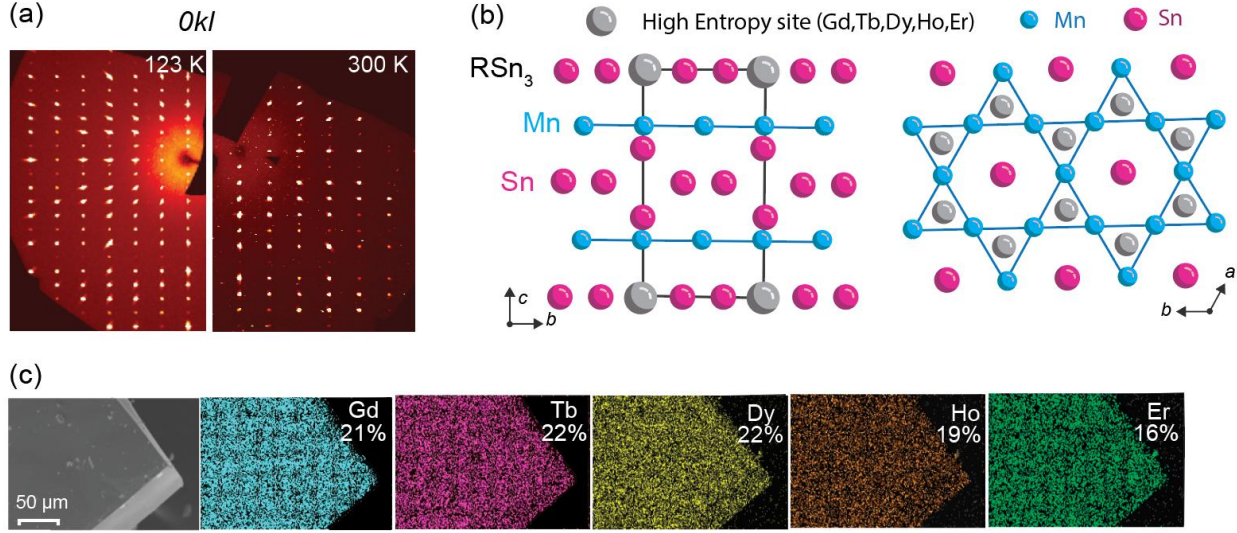

Fig. 1, Structure and composition of HE166. a) Precession image of HE166 obtained from X-ray diffractometer at room temperature at 300K and low temperature at 123 K. b) Crystal structure of HE166 viewed along the crystallographic *a* and *c* axis. c) SEM-EDX mapping revealed the chemical distribution of the rare earth elements in HE166, highlighting the homogeneous distribution of the five rare-earth elements.

Table 1: Crystallographic data of HEMn$_6$Sn$_6$ at 123 K and 300 K.

| Compounds | (Gd$_{0.21}$,Tb$_{0.22}$,Dy$_{0.22}$,Ho$_{0.19}$,Er$_{0.16}$)Mn$_6$Sn$_6$ | |
|---|---|---|
| Temperature | 123 K | 300 K |
| Crystal symmetry | hexagonal | hexagonal |
| Space Group | P6/mmm (No.191) | P6/mmm (No.191) |
| Lattices | $a = b = 5.5056(3)$ Å, $c = 8.9869(7)$ Å $V = 235.91(3)$ Å$^3$, Z = 1 | $a = b = 5.5273(2)$ Å, $c = 9.0118(6)$ Å $V = 238.43(2)$ Å$^3$, Z = 1 |
| Absorption coefficient | 31.070 mm$^{-1}$ | 30.741 mm$^{-1}$ |
| F(000) | 516.0 | 516.0 |
| R$_{int}$ | 0.022 | 0.026 |
| Reflections collected | 5081 | 4016 |
| Independent reflections | 233 | 235 |
| Extinction coefficient | 0.0199(15) | 0.061(4) |
| Data / restraints / parameters | 233 / 1 / 20 | 235/1/20 |
| Goodness-of-fit on F$^2$ | 1.316 | 1.362 |
| Refine_diff_density_max | 1.567 | 2.636 |
| Refine_diff_density_min | -2.083 | -3.533 |
| R$_1$, wR$_2$ [I > 2 σ] | 0.0192, 0.0529 | 0.0307, 0.0686 |
| R$_1$, wR$_2$ (all data) | 0.0192, 0.0529 | 0.0307, 0.0686 |

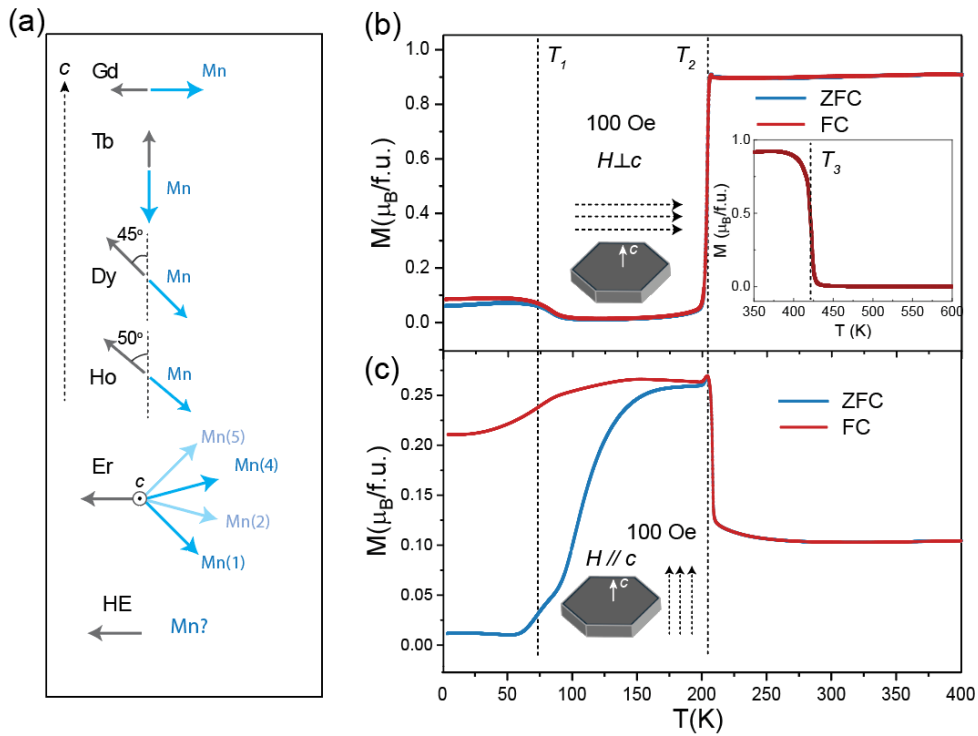

Fig. 2 a). Magnetic order of the parent phase with different rare earth elements. b) c) Temperature dependence of magnetization on HE166 with external magnetic field applied along and perpendicular to the *c* axis.

The 166 parent phases exhibit different magnetic structures depending on the rare earth element,[5,10,13,17–19] which is illustrated in Fig. 2a. The light blue and grey arrows indicate the magnetic orders of Mn and rare earth elements respectively. In the high-entropy 166 system, where all the rare earth elements occupy the same crystallographic site, no evidence of phase separation was observed in our single crystal diffraction, chemical analysis as well as the magnetization measurements. Furthermore, the magnetic phases in the high-entropy 166 do not appear to be a superposition of parent phases containing only one rare earth element. Instead, the competition among the different magnetic ions from Gd to Er leads to a richer magnetic phase diagram including new transitions as well as new transition temperatures that do not appear in the parent phases. Fig. 2b and 2c show the temperature dependent magnetizations of HE166 with magnetic field applied along the in-plane and out-of-plane directions, respectively. Multiple magnetic transitions have been observed, which are marked as $T_1$, $T_2$ and $T_3$. Fig. 2b shows the temperature dependence of the magnetization when an external magnetic field is applied within the basal plane. Upon cooling from 600 K, a sudden increase of the magnetization occurs at 420 K, marked as $T_3$, indicating the paramagnetic to ferrimagnetic phase transition, as shown in the inset of Fig. 2b. As temperature continues to decrease to 200 K, the magnetization remains nearly constant, indicating the presence of a saturated in-plane ferrimagnetic order. Below 200 K, the magnetization drops dramatically, indicating that a spin reorientation occurs where the in-plane ferrimagnetism changes to out-of-plane uniaxial ferrimagnetism, which we will discuss further later. Below 80 K, a slight splitting occurs between the zero-field cooling (ZFC) and field cooling (FC) magnetization, and this is related to another magnetic transition associated with the spin reorientation, as seen in other R166 systems.[20,21] Fig. 2c shows the magnetization data, when the external magnetic field is applied out-of-plane. A peak occurs at 200 K, corresponding to the spin reorientation. Slightly below 200 K, the splitting between ZFC and FC magnetization occurs. When the temperature continuously decreases, a small hump feature as well as kink-like behavior happen below 80 K, consistent with the in-plane measurements, jointly revealing the spin-reorientations transition.

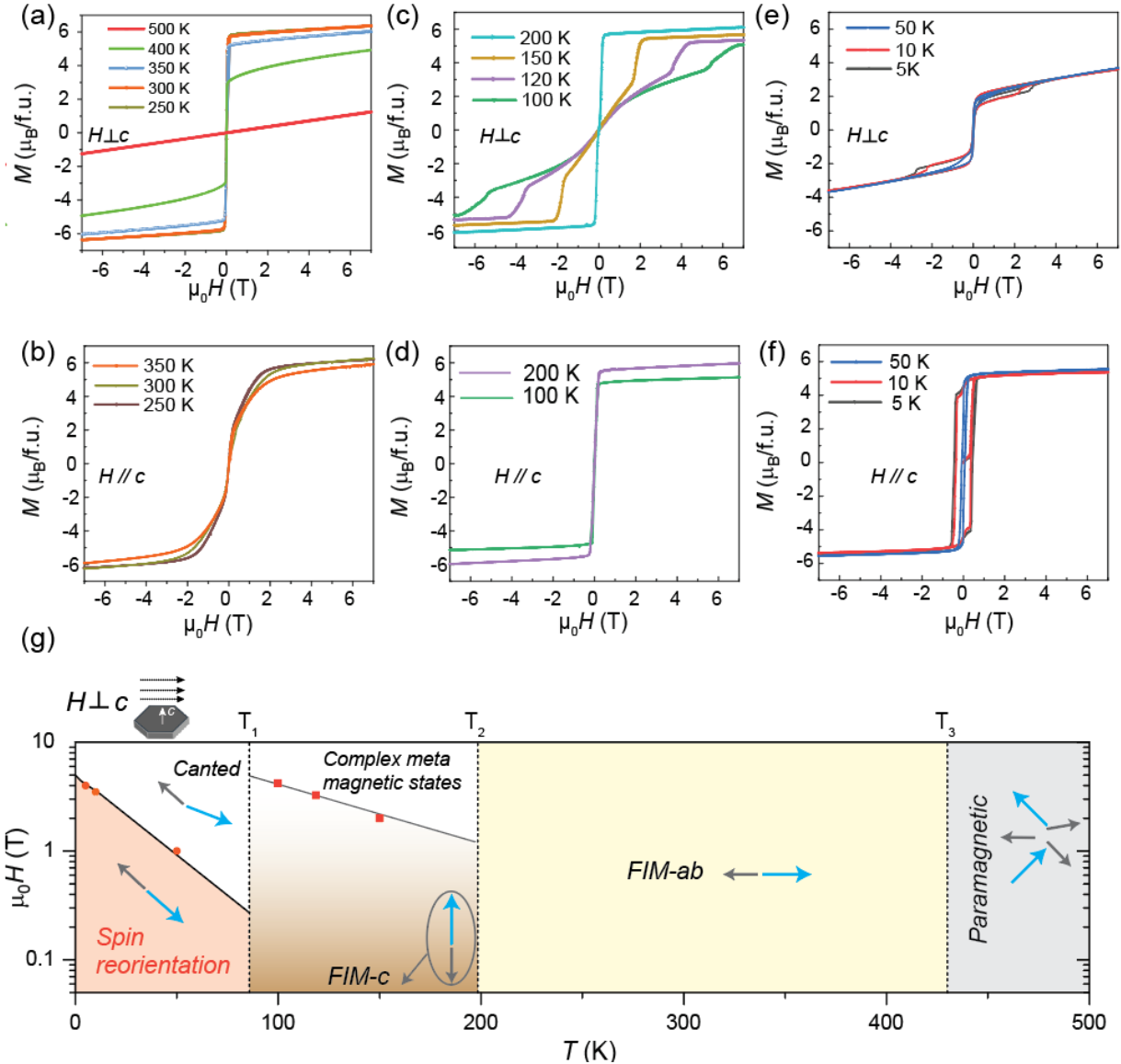

Fig. 3. Field-dependent magnetization hysteresis loops of HE166 at various temperatures. a),c),e) Measurements from 5-350 K with external magnetic applied along in-plane direction (H ⊥ c). b),d),f) Measurements from 5-500 K with field applied along out-of-plane direction (H // c). g) Magnetic phase diagram of HE166 with external magnetic field applied perpendicular to crystalline *c* direction. *FIM-ab*: in plane ferrimagnetism. *FIM-c*: uniaxial ferrimagnetism. Light blue and grey arrows: magnetic orders of high-entropy rare-earth elements and Mn atoms, respectively.

To further understand the magnetic interactions, magnetic hysteresis loops at fixed temperatures were measured, as shown in Fig. 3. These measurements reveal large anisotropies and rich magnetic behaviors in this HE166 system. When the external magnetic field is applied in-plane ( H ⊥ c ) , the M-H curve at 500 K shows a linear response, confirming paramagnetic behavior of HE166, as shown in Fig. 3a. Upon cooling to 400 K, the magnetization undergoes a sharp increase and can reach saturation easily under small external field. This saturation becomes more pronounced with further cooling. Similar magnetization behavior of HE166 is observed in the temperature range of 200- 100 K when the magnetic field is applied out-of-plane (H //c), as shown in Fig. 3e. They jointly indicate a reorientation of the ferrimagnetic easy axis from the in-plane to out-of-plane direction at 210 K, as revealed in the M (T) measurement in Fig. 2.

The field induced magnetization exhibits dramatically different behavior when the external field is applied perpendicular to the easy axis of ferrimagnetic order at different temperatures, as seen in Fig. 3b and 3c. In Fig. 3b (H // c), themagnetization between 350 – 250 K increases gradually before reaching saturation, indicating a continuous spin reorientation as the ferrimagnetic order is forced to align with the external field. In contrast, Fig. 3c (H ⊥ c) shows that metastable magnetic states are easily induced between 200 and 100 K when the external magnetic field is perpendicular to the ferrimagnetic order. For example, at 150 K, the magnetization increases gradually until a sharp jump at 2 T, followed by a slow increase towards saturation above 3 T. Such an unusual magnetization resembles the topological Hall effect, which has been revealed in $ErMn_6Sn_6$[10].

Below 50 K, both the magnetization curves with H // c and H ⊥ c share common features of hysteresis loops, which become more prominent upon cooling, as seen in Fig. 3e and 3f. The formation of hysteresis loops indicates the deviation of the easy axis of FIM order from the *c* axis, as shown in the Dy, Ho and Er 166 parent systems[14,17]. It should be noted that magnetization does not saturate with H ⊥ c, as seen in Fig. 3e, suggesting the formation of spin canting where the magnetic moments of the Mn and rare earth elements tilt away from a collinear alignment.

Considering all the magnetization results, a magnetic phase diagram is constructed with the external magnetic field applied perpendicular to the crystalline *c* direction (*H* ⊥ *c*), as shown in Fig. 3g. The competition between the multiple factors like exchange, anisotropy and Zeeman energies[10] yields a rich magnetic phase diagram. Above $T_3$, thermal spin fluctuations dominate and the whole system shows paramagnetic behavior. Below $T_3$ and above $T_2$, exchange energy prevails, stabilizing a long-range planar ferrimagnetic order. When the temperature further decreases below $T_2$ but above $T_1$, this ferrimagnetic order reorients from in-plane direction to the out-of-plane direction, due to the dominance of the out-of- plane anisotropy energy. Notably, within this intermediate temperature range from 80 to 200 K, the out-of-plane magnetic state exhibits complex metamagnetic behavior under in-plane magnetic fields, with *M–H* loops suggestive of non-zero scalar spin chirality and potentially chiral spin textures[7,8,12,22]. Verification of such states requires advanced neutron diffraction studies. Below $T_1$, thermal fluctuations are further suppressed, and the anisotropy energy increasingly governs the spin configuration, inducing an additional spin reorientation in which the FIM easy axis tilts away the *c* axis towards an intermediate direction between the *c* axis and the *ab* plane. Under external magnetic fields, the Mn and rare-earth moments develop slightly canting, forming canted spin textures.

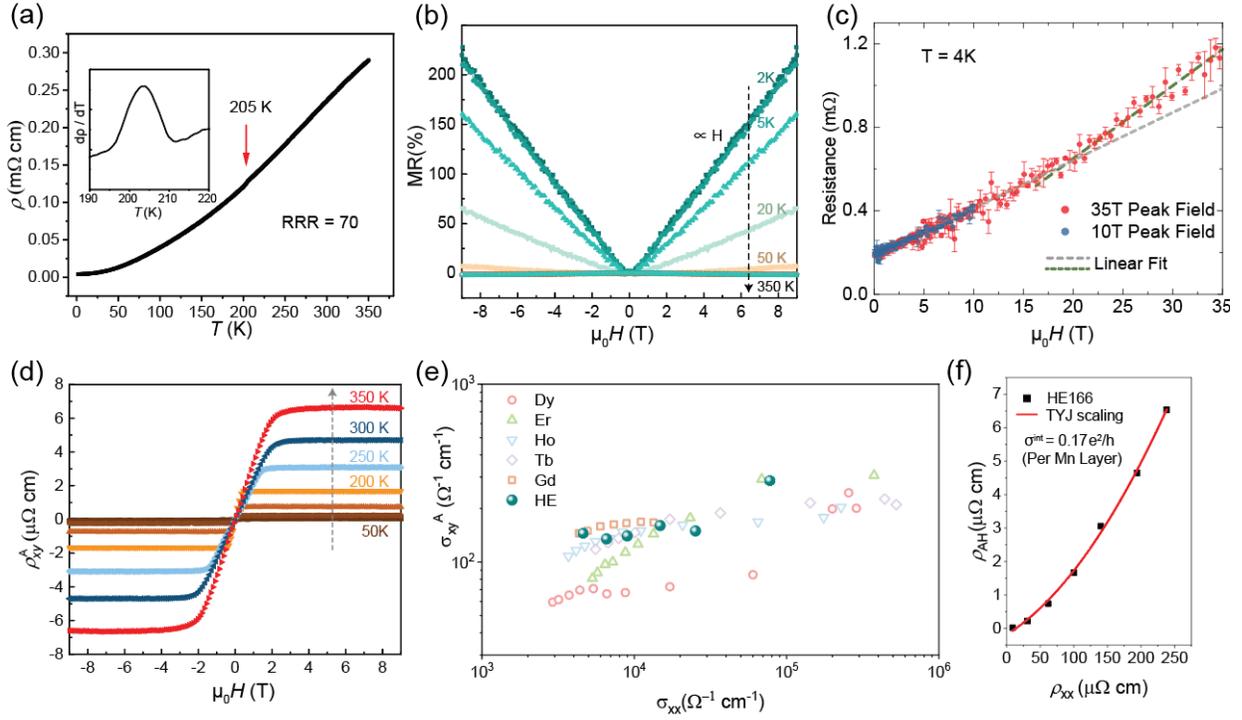

Fig.4. Electrical transport results for HE166. a) Temperature dependent resistivity of HE166 system. b) Magnetoresistance of HE166 from 0 – 9 T at various temperatures from 4 - 350 K. c) Magnetoresistance of HE166 at 4 K in pulsed magnetic fields up to 35 T. d) Hall resistivity of HE166 as a function of magnetic field at various temperatures. e) Anomalous Hall conductivity vs. longitudinal conductivity of HE166 (solid balls) and the parent phases (open circles). f) Tian-Ye-Jin scaling of HE166.

Besides magnetization measurements, electrical transport measurements were also carried out to understand the band topology in the high entropy kagome lattice HE166. In-plane resistivity as a function of the temperature and magnetic field is shown in Fig. 4. The ρ(T) indicates the metallic nature of the sample over the whole temperature range of 2 - 350 K, which is similar to the parent phases. The residual resistance ratio (RRR) defined as RRR = $R_{xx}(300K) / R_{xx}(2K)$] is about 70, indicating the high quality of our crystals despite the additional electron scattering caused by high entropy alloying. The derivative of resistivity shows a small yet clear peak at 205 K, as shown in the Fig. 4a inset, which is correlated with the magnetic transition $T_2$ discussed previously. Magnetoresistance $MR = [\rho_{xx}(H) - \rho_{xx}(0)]/\rho_{xx}(0) \times 100\%$ is shown on Fig. 4b. A clear non-saturating linear MR with a ratio of 225% at 2 K is observed. Compared to the parent phases, HE166 with five rare earth elements shows enhanced linear behavior, which may arise from the high entropy[18]. As the temperature is increasing, the MR keeps its linearity with a reduced MRR ratio.

To explore further the extent of the linear MR in HE166, the MR was measured at T = 4 K under higher magnetic fields in a pulsed magnet with peak fields of 10 and 35 T, as shown in Fig. 4c. The linearity remains until the magnetic field reaches 20 T, where a kink occurs and is then followed by another relatively linear MR regime. There is no signs for saturation for the second linear MR behavior up to 35 T. The kink around 20 T has never been reported in the parent phases of Gd, Tb, Dy, Ho166 systems. In the case of

ErMn$_6$Sn$_6$, which is close to the boundary of spiral magnetism and ferrimagnetism, several theoretical works report a FIM to canned magnetic structures occur for all the parent phases around 20T [10,23]. The change of the coupling between Er and Mn around 20 T seems a plausive reason for the kink in the HE166 system here.

Furthermore, the anomalous Hall effect can reveal subtle interactions between localized moments and itinerant electrons that are not captured by conventional transport measurements. The evolution of anomalous Hall resistivity with temperature can help distinguish contributions from intrinsic mechanisms and extrinsic scattering process. The former is generally related with band topology and that later are involved with the impurity induced side jump and skew scattering[24–27]. The anomalous Hall resistivity is separated via the empirical relation,

$$\rho_{xy} = \rho_{xy}^0 + \rho_{xy}^A = R_0 B + 4\pi R_s M$$

Here the first term $\rho_{xy}^0$ indicates the normal Hall resistivity and the second term represents the anomalous Hall effect, which is correlated with the magnetic order in the system. We focus just on the second term. When the temperature increases, the saturation value of $\rho_{xy}$ becomes bigger. However, in the low temperature region below 50 K, the noise is too large due to the small signal, which we will not take into consideration. The conductivity tensors of $\sigma_{xx}$ and $\sigma_{xy}$ are derived by $\sigma_{xx} = \rho_{xx}/(\rho_{xy}^2 + \rho_{xx}^2)$ and $\sigma_{xy} = \rho_{xy}/(\rho_{xy}^2 + \rho_{xx}^2)$. The scaling behavior between $\sigma_{xy}$ and $\sigma_{xx}$ of HE166 and the parent phases is shown in Fig. 4e. HE166 shows similar trends to the parent phases. All the $\sigma_{xy}$ are weakly linked to $\sigma_{xx}$ and clearly deviate from a linear relation, a characteristic of an extrinsic mechanism like screw scattering and side jump instead of an origin from disorder or impurities.[24] The deviations of linear relation indicate the contributions from the intrinsic term, which depends only on the band structure and indicates the existence of nontrivial band structures like Berry curvatures. To extract the intrinsic Hall conductivity, the Tian-Ye-Jin scaling law is utilized[28].

$$\rho_{xy}^A = \sigma_{int}\rho_{xx}^2 + \sigma_{skew}\rho_{xx}$$

$\sigma_{int}$ is estimated to be 66 Ω$^{-1}$cm$^{-1}$. For each Mn layer, $\sigma_{int}$ is estimated to be 0.17 e$^2$/h. The parent phases show $\sigma_{int}$ ranges from 0.04 – 0.27 e$^2$/h per Mn Layer.[6,14,18,29] As a result, the estimated $\sigma_{int}$ in the high entropy phase is close to the average value of the parent phases,[14] providing another piece of evidence for the modulation of the topological phases in R166 system through high entropy engineering.

In summary, we have synthesized a high entropy kagome lattice with the five rare earth elements of Gd, Tb, Dy, Ho and Er. The introduction of high compositional disorder into the kagome system results high entropy, which leads to a series of emergent phenomena not observed in the parent compounds. Furthermore, a non-saturating magnetoresistance with enhanced linearity is observed in HE166. An intrinsic anomalous Hall effect is also demonstrated, indicating the survival of non-trivial band topology in HE166. Our work supports high- entropy as a powerful approach for tuning the interplay of charge, spin and lattice in magnetic topological materials.

**Material synthesis:**

High entropy 166 single crystals were synthesized using a flux method where Sn is used as the flux. High-purity Gd (Alfa Aesar, 99%), Tb (Alfa Aesar, 99.9%), Dy (Alfa Aesar, 99.8%), Ho (Alfa Aesar, 99.9%), Er (Alfa Aesar, 99.9%) were filed from ingots, then mixed with Mn (Alfa Alser, 99.99%) pieces and Sn (Alfa Aesar 99.9999%) ingots with the ratio of 0.2:0.2:0.2:0.2:0.2:6:20. The mixtures were placed in an alumina crucible and subsequently sealed in an evacuated quartz tube. The assembly was heated up to 1100 °C in a furnace and held at this temperature for two days, which is followed by cooling down to 600°C at a rate of 1 °C/h. After the tube was centrifuged at 600°C, large size single crystals with an average lateral size of 5 x 5 mm$^2$ crystals was obtained.

The chemical composition of the yield crystals was verified by energy-dispersive x-ray spectroscopy (EDX) on a DM07 Zeiss Supra 40 scanning electron microscope. The crystal structure of the HE166 phase was determined by single-crystal X-ray data measured on a Bruker DUO diffractometer equipped with an Apex II area detector and an Oxford Cryosystems 700 Series temperature controller with a Mo $K\alpha$ source ($\lambda = 0.7103$Å). The collected data set was integrated using the Bruker Apex-II program, with the intensities corrected for the Lorentz factor, polarization, air absorption, and absorption due to variation in the path length through the detector faceplate. The data were scaled, and absorption correction was applied using SADABS. The structure was solved by using the intrinsic phasing method in SHELXT and refined using SHELXL with all atoms refined anisotropically.

**Magnetization and transport measurement:**

DC magnetization was conducted with a Quantum Design Magnetic Property Measurement System (MPMS3) down to 5 K. Resistivity and Hall signals of HE166 single crystal were measured using a Quantum Design Physical Property Measurement System (PPMS) with the temperature range of 5 to 300 K and the magnetic fields up to 9 Tesla. Measurements in pulsed magnets were conducted at the National High Magnetic Field Laboratory's Pulsed Field Facility in Los Alamos National Laboratory. Four-wire electrical resistance measurements were performed in a 65T short pulse magnet with a He-4 cryostat. Owing to the small resistance of the sample, resistance was measured with a pulsed direct current method based upon a modified form of the approach described in Ref. [30]. Typical current durations were on ~10$\mu$s. Each data set is the average of multiple pulses to improve the signal-to-noise ratio, and the resistances were symmetrized with respect to the direction of the applied magnetic field. The maximum applied magnetic field was limited by the strong torque response of the sample.

**Acknowledgements:**

The work conducted at University of Texas at Dallas was supported by US Air Force Office of Scientific Research Grant No. FA9550-19-1-0037, National Science Foundation- DMREF-2324033, and Office of Naval Research grant no. N00014-23-1-2020. A portion of this work was performed at the National High Magnetic Field Laboratory, which is supported by the National Science Foundation Cooperative Agreement No. DMR-2128556, the State of Florida, and the Department of Energy. Work at the University of California, Berkeley and Lawrence Berkeley National Laboratory was funded by the U.S. DOE, Office of Science, Office of Basic Energy Sciences, Materials Sciences and Engineering Division under Contract No. DE-AC02-05CH11231 (Quantum Materials Program KC2202).